\documentclass[a4paper,12pt]{spieman}  
\usepackage{amsmath,amsfonts,amssymb}
\usepackage{siunitx}
\usepackage{graphicx}
\graphicspath{{figs}{}}
\hypersetup{pdfstartview={FitH},pdfpagemode={UseNone},
            colorlinks,linkcolor=blue, citecolor=blue, urlcolor=blue,
            bookmarksopen=true, pdfnewwindow=true}
\usepackage[all]{hypcap}

\usepackage{setspace}
\usepackage{tocloft}
\usepackage{lineno}
\usepackage{physics}
\usepackage{newtxmath}
\title{Inverse designed metasurfaces for multi-wavelength full-Stokes polarisation imaging}

\author[a,b]{Sarah~E.~Dean*}
\author[a]{Neuton Li}
\author[a]{Josephine Munro}
\author[a,c]{Benjamin Laudert}
\author[b,c,d]{Thomas Siefke}
\author[b,c]{Gia Quyet Ngo}
\author[e]{Robert Sharp}
\author[a]{Dragomir N. Neshev}
\author[b,c]{Falk Eilenberger}
\author[a]{Andrey~A.~Sukhorukov}
\affil[a]{ARC Centre of Excellence for Transformative Meta-Optical Systems (TMOS), Department of Electronic Materials Engineering, Research School of Physics, The Australian National University, Canberra, ACT 2600, Australia}
\affil[b]{Fraunhofer-Institute for Applied Optics and Precision Engineering IOF, Albert-Einstein-Str. 7,
07745 Jena, Germany}

\affil[c]{Friedrich Schiller University Jena, Abbe Center of Photonics, Institute of Applied Physics, Albert-Einstein-Str. 15, 07745 Jena, Germany}

\affil[d]{Ernst-Abbe-Hochschule Jena University of Applied Sciences, Carl-Zeiss-Promenade 2, 07745 Jena, Germany}
\affil[e]{Research School of Astronomy \& Astrophysics, The Australian National University, Canberra, ACT 2611, Australia}

\cftpagenumbersoff{figure}
\cftpagenumbersoff{table} 
\begin{document} 
\sloppy
\maketitle

\begin{abstract}
Multispectral full-Stokes polarisation imaging has a broad range of applications, from biological cell imaging to agricultural remote surveying. For such applications, especially involving lightweight unmanned aerial vehicles like drones, it is necessary to have compact, single-shot, efficient optical systems.
We present a topology-optimised metasurface design that diffractively separates a scene into spectral and polarimetric measurements, operating for 532~\unit{\nano\metre} and 700~\unit{\nano\metre} in a single-shot imaging system. The polarisation imaging performance of the design is shown in simulation to be robust to critical performance metrics, matching both spectral and angular bandwidth requirements. We fabricated our metasurface design by nanopatterning a TiO\textsubscript{2} thin film, and experimentally demonstrate polarisation reconstruction for two wavelengths with a single metasurface structure.

\end{abstract}

{\noindent \footnotesize\textbf{*}Corresponding author: Sarah~E.~Dean, \linkable{sarah.dean@anu.edu.au} }

\begin{spacing}{1.5}  

\section{Introduction}
\label{sec:intro}  

Polarisation imaging is the process of measuring the orientation of light waves across a scene of interest to gain additional insights beyond what is achievable with intensity imaging. A broad range of applications benefit from the addition of polarisation imaging, including aerosol monitoring\cite{Tyo2006, NASA-PACE,Sparks2009}, biological cell imaging\cite{Thrane2026,Li2025b}, or agricultural applications where the polarisation response at multiple wavelengths can provide beneficial information for crop identification or vegetation health monitoring\cite{Patty2018,MaraisSicre2020,Kudenov2023,Zhu2025}.

Since conventional visible light and shortwave infrared detectors are not inherently sensitive to polarisation, additional polarisation-selective optical components are required. Typically, polarisation measurements are performed using either arrays of filters, prisms, or combinations of polarisers and rotating waveplates\cite{Sabatke2000,Rubin2019}. While these methods are well-established and robust, each has associated limitations with regards to applications that require lightweight, compact, or single-shot systems.

For multispectral polarimetry, the limitations of these conventional methods are amplified; each targeted wavelength requires additional beamsplitting or filtering to capture measurements, increasing the size of the system or decreasing photon flux. Therefore, for applications where constraints with respect to size, weight, or available light exist, it is necessary to utilise non-conventional polarisation imaging methods.

\begin{figure}[h]
\centering
  \includegraphics[width=\columnwidth]{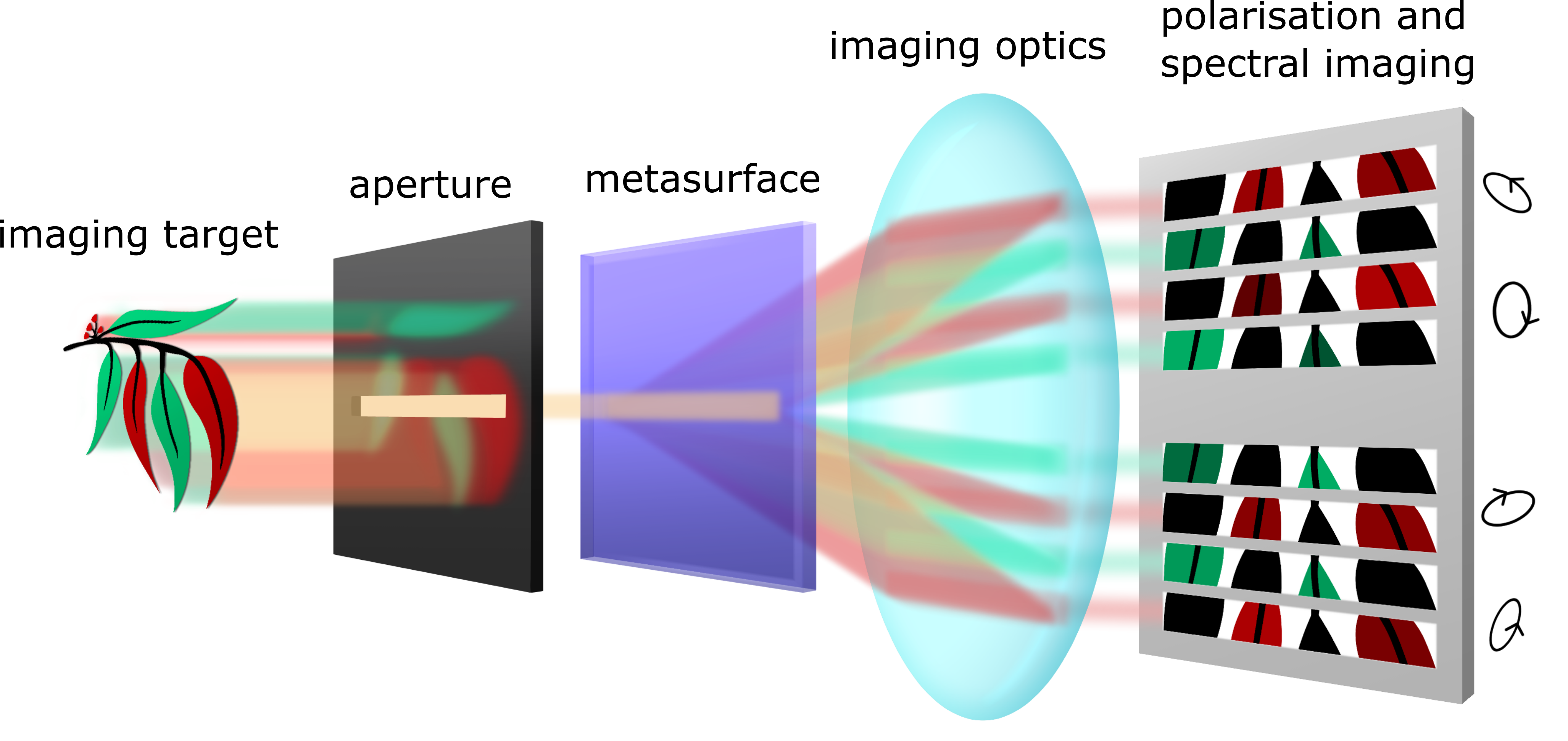}
  \caption{Illustration of the proposed metasurface behaviour, demonstrating the concept of mixed wavelength incident light being diffracted into distinct spectral and polarisation measurements for wavelength-specific single-shot polarimetry.}
  \label{fig:conceptual}
\end{figure}

Due to the accessibility of drones and other lightweight unmanned aerial vehicles for surveying applications, it is attractive to develop compact single-shot polarisation imaging for such platforms. One existing field of research producing compact devices is metasurface optics, which consist of a nanostructured array of metallic or dielectric optical resonators on a flat substrate. The material and shapes of the resonators produce a tailored optical response\cite{Kivshar2017}, allowing metasurfaces to be used for many applications, such as prisms, lenses, waveplates, and multifunctional combinations of optical elements\cite{Arbabi2017,He2018,Shi2020,Wang2018}. 

There are also many examples of single-shot polarisation imaging achieved with metasurfaces\cite{Arbabi2018,Fan2023,Zuo2023,Rubin2019,Huang2023,Ren2023,Dean2025, Li2026}; however, previous works focus on monochromatic incident light, limiting the scope of applications of any one device. Works implementing multispectral polarisation imaging do so with multiple metasurface structures of different functionalities\cite{Li2023, Hu2024}, inciting the question on if we can further integrate the necessary functionality for polarisation imaging within a single metasurface structure into a smaller optical footprint. 

We address this problem by developing a diffractive metasurface designed for multispectral polarimetry at $532$~\unit{\nano\metre} and $700$~\unit{\nano\metre} as illustrated in Fig.~\ref{fig:conceptual}, using diffraction as an intrinsic method of spectrally separating our measurements.  We also simulate critical performance metrics such as operational field-of-view and spectral bandwidth, using  potential agricultural remote surveying applications as a basis for our operating specifications and performance. Finally, we also present preliminary experimental characterisation of initial metasurface samples, demonstrating accurate polarisation reconstruction at two separate wavelengths.

\section{System Design }
\subsection{Field-of-view considerations}

To ensure our multispectral imaging measurements do not overlap between diffraction orders, and to take advantage of the movement of a drone-based system, we base our system around the push-broom imaging technique. This is an established technique in remote sensing and surveying, where a narrow image or \textit{swath} is continuously imaged as the system translates perpendicular to the swath major axis. A full image is then created by stitching together adjacent images into a continuous two dimensional scene\cite{Gupta1997,Koduri2012}.

Push-broom imaging is often used in conventional multispectral imaging using a division-of-aperture technique; the imaging swath is separated into different wavelength channels, typically using a one-dimensional diffraction grating, and the channels are then imaged onto a two-dimensional detector. This simultaneous imaging at different wavelengths avoids errors such as illumination changing between measurements, imaging offsets between measurements, or movement within the scene. We utilise this same division-of-aperture principles for our multispectral polarimetry design to avoid these unwanted effects, using a metasurface diffracting in one direction aligned with the narrow imaging swath to naturally separate our wavelengths into distinct spectral channels. Here the individual diffraction orders act as separate polarisation measurement channels, allowing us to record all polarisation and spectral measurements simultaneously, as illustrated in Fig.~\ref{fig:basicresults}(a).

\subsection{Polarimetry measurements \label{sec:pol}}
Full-Stokes polarimetry allows the measurement of the full polarisation state of light, including elliptical and partially polarised states, requiring a minimum of four polarisation-selective intensity measurements. The optimum measurements for a four-measurement polarimeter are well-established in literature\cite{Azzam1988,Sabatke2000, Sabatke2000a,Ambirajan1995, Foreman2015}, and have been used as the analysis states in several previous metasurface polarimetry works\cite{Rubin2019,Rubin2022,Li2023,Dean2025}. 

The Stokes vector representation of light is given by $\Vec{S}=[S_0,\;S_1,\;S_2,\;S_3]^T$, where $S_0$ is the total intensity of light, the $S_1,\;S_2,\;S_3$ components define the polarisation state. For an incident Stokes polarisation state $\Vec{S}$, and a set of polarisation-selective intensity measurements $\Vec{I}=[I_{-2},\;I_{-1},\;I_1,\;I_2]^T$, corresponding to the diffractively separated measurements as described in Section \ref{sec:intro}, the relationship between the input polarisation and output measurements can be described by an instrument matrix $\mathbf{M}$ such that:
\begin{equation}
    \mathbf{M}\Vec{S}=\Vec{I} \,.
    \label{eq:MMat}
\end{equation}
The precise instrument matrix of a polarimetry system is calculated from the relationship between a set of calibration polarisation input states and the respective output intensity measurements, ensuring the polarimetry system is robust to variance between the designed and implemented polarimetry measurements. The calibrated instrument matrix is then used to reconstruct unknown input polarisations, with the maximum likelihood method utilised for additional robustness to measurement error in the calculation\cite{James2001,Lung2024}.

The maximum likelihood method is a technique that numerically optimises a density matrix to perform the reconstruction of polarisation states, ensuring that the resulting reconstruction is robust to any measurement errors in the system, and that any results are strictly physical. The maximum likelihood method used in this work is a classical implementation of the process by James et al. \cite{James2001}, which has been used in other works \cite{Lung2024}, and is detailed here for our polarisation reconstruction.

To ensure that our density matrix is explicitly physical, we define the general form as:
\begin{equation} \label{eq:rhoT}
    {\rho}=\mathbf{T}^\dagger\mathbf{T}/\mathrm{Tr}\{\mathbf{T}^\dagger\mathbf{T}\}, \text{where}\;\mathbf{T}=\mqty[t_{1}&&0\\t_{2}+it_3&&t_{4}].
\end{equation}
The density matrix can be related to the Stokes parameters of our polarisation reconstruction as\cite{Lung2024}:
\begin{equation} \label{eq:rhoS}
    {\rho}=\frac{1}{2}\sum_{j=0}^3 S_j\sigma_j=\frac{1}{2}\mqty[S_0+S_3 && S_1-iS_2 \\ S_1+iS_2 && S_0-S_3],
\end{equation}
where $\sigma_j$ are the Pauli matrices. By inverting Eq.~(\ref{eq:rhoS}) and using Eq.~(\ref{eq:rhoT}), we define a strictly physical reconstructed polarisation state $\Vec{S}_\mathrm{R}$ through $(t_{1},t_{2},t_{3},t_{4})$.
We perform the reconstruction by numerically minimising the following function, corresponding to a least-squares solution of the associated set of equations:
\begin{equation}
    f(t_{1},t_{2},t_{3},t_{4})=\left\lVert\;\mathbf{M}\Vec{S}_\mathrm{R}(t_{1},t_{2},t_{3},t_{4}) - \Vec{I}\;\right\rVert^2,
\end{equation}
where $\mathbf{M}$ is the calibrated instrument matrix, and $\Vec{I}$ is the polarisation-selective intensity measurements for an unknown polarisation.

\section{Metasurface design framework\label{sec:framework}}

There are two key components determining the design parameters of the metasurface; selection of the wavelengths for the multispectral polarimetry, and determining the structural parameters and design method suitable for a polarimetry metasurface operating at the chosen wavelengths.

The choice of wavelengths are based on both the target application and the diffraction angles of each wavelength. To guide our work with a practical motivation, we target the chlorophyll reflection and absorption spectral bands\cite{Li2025} to enable the potential agricultural applications described earlier. More critically, our chosen wavelengths must not have overlapping diffraction orders, such that there is a clearly separable signal for the $\pm 1$ and $\pm 2$ diffraction orders.
Ensuring the diffractive process alone separates the wavelengths helps to maintain a compact and lightweight imaging system by eliminating the need to use additional optics to separate the incident wavelengths. Therefore, we design our metasurface to image at $532$~\unit{\nano\metre} and $700$~\unit{\nano\metre} wavelengths, which are within the target chlorophyll reflection and absorption bands and have well-separated diffraction orders.

Suitable materials and structures of the metasurface area determined by the operating wavelength and desired behaviour of the metasurface. We consider a structure of 300~\unit{\nano\metre} thick patterned titanium dioxide (TiO\textsubscript{2}) on a 500~\unit{\micro\metre} substrate, as TiO\textsubscript{2} has a high refractive index that enables light to resonate within and be manipulated by the metasurface\cite{Yang2017}. TiO\textsubscript{2} is also transparent within the visible spectrum, ensuring transmission losses do not stem from material absorption. The thickness of the metasurface affects the deflection efficiency of light to larger angles\cite{Yang2017}; we determined the thickness of our patterned layer by comparing the performance of designs created with different thicknesses, and finding the highest performing metasurface design within the fabrication capabilities available to us has a thickness of 300~\unit{\nano\metre}.

We choose a periodicity of $4.5$~\unit{\micro\metre}~$\times$~$0.25$~\unit{\micro\metre} for our patterned metasurface layer, as it produces one-dimensional diffraction with orders at $\pm 8.9^\circ,\;\pm 18.1^\circ$ for $700$~nm and $\pm 6.8^\circ,\;\pm 13.7^\circ$ for $532$~nm incident light, angularly separating the spectral components sufficiently for imaging. The diffraction orders of interest are also propagating with a small enough angle from normal such that we can use an objective with a minimum numerical aperture of $\mathrm{NA} = 0.32$ to capture the analysis diffraction orders, allowing the system surrounding the bespoke metasurface to operate with simple optics.

To design the patterned TiO\textsubscript{2} layer, we utilise topology optimisation. This design technique enables more complex and efficient metasurface behaviour due to the freeform structures that can form during the optimisation process. Furthermore, topology optimisation facilitates better control of the transmission to the unused centre diffraction order, and allows us to optimise for both wavelengths simultaneously within the same metasurface structure\cite{Sell2017,Fan2020, LalauKeraly2013,Jensen2011}.

Consistent with previous topology optimisation metasurface works\cite{Sell2017,Fan2020, LalauKeraly2013,Jensen2011,Dean2025}, our optimisation process begins with an array of smooth, random values between the refractive index of air and of our metasurface material, forming the design space of a single metasurface period. For each iteration of the optimisation, we perform a forward and adjoint simulation, and then use Lorentz reciprocity to calculate the gradient in transmission as a function of refractive index. We perform a gradient descent operation to optimise the design towards the carefully crafted figure-of-merit (FoM) described further in this section, with binarisation and robustness functions each iteration, until the optimisation converges to a discrete final design that satisfies the feature size requirements for the fabrication methods and systems used later in this work. 

We performed our simulations using rigorous coupled-wave analysis in the MATLAB package RETICOLO\cite{Hugonin2022}. The instrument matrix at each wavelength was calculated at each iteration, and the polarisation reconstruction performance of the measurements at both wavelengths are maximised concurrently. One established figure-of-merit for polarimetry is the condition number, corresponding to the amplification of errors in the system during reconstruction\cite{Azzam1988,Ambirajan1995, Sabatke2000,Sabatke2000a, Peinado2010, Foreman2015}. The condition number is calculated from the singular values $\mu_{\lambda,i}$ of the instrument matrix at one wavelength $M_\lambda$. The singular values of $M_\lambda$ are given by the singular value decomposition:
\begin{equation}
    \mathbf{M}_\lambda = \mathbf{U}_\lambda\mqty[\mu_{\lambda,1} && 0 && 0 && 0 \\
    0 && \mu_{\lambda,2} && 0 && 0 \\
    0 && 0 && \mu_{\lambda,3} && 0 \\ 
    0 && 0 && 0 && \mu_{\lambda,4}]\mathbf{V}_\lambda^\dagger ,
\end{equation}
where $\mathbf{U}_\lambda$ and $\mathbf{V}_\lambda$ are 
unitary matrices. The condition number is calculated from this decomposition as:
\begin{equation}
    \mathrm{cond}(\mathbf{M}_\lambda)=\frac{\max_{i}(\mu_{\lambda,i})}{\min_{i}(\mu_{\lambda,i})}.
\end{equation}
The minimum condition number of optimum four-measurements system established in previous works\cite{Azzam1988,Ambirajan1995, Sabatke2000,Sabatke2000a, Peinado2010, Foreman2015} is $1.7$, indicating the four measurements perfectly span the Stokes space and corresponding to an error amplification factor of $1.7$. 

As the maximum singular value is bounded by transmission, and we also want to optimise for greater transmission, we choose to maximise the minimum singular value for our optimisation figure-of-merit ($\mathrm{FoM}$):
\begin{equation}
  \mathrm{FoM}_\lambda=\min_{i}(\mu_{\lambda,i}).
  \label{eq:FoMpart}
\end{equation}
To form the $\mathrm{FoM}$ for simultaneously optimising the polarisation reconstruction performance at multiple wavelengths, we take the product of Eq.~(\ref{eq:FoMpart}) for each incident wavelength: 
\begin{equation}
    \mathrm{FoM}=\min_{i}(\mu_{532\mathrm{nm},i})\cdot\min_{j}(\mu_{700\mathrm{nm},j}).
\end{equation}

We optimise for the overall polarisation reconstruction performance instead of optimising for a set of specific polarisation measurements for two key reasons. Firstly, the performance of the optimal set of analysis states is invariant to the global rotation of these states in the Stokes space; the key property is the span of the states relative to each other. Secondly, the increased complexity of the system means identical polarisation responses may not be achievable with the same structure for both wavelengths simultaneously. Instead, allowing the polarisation response to be different for both wavelengths allows the overall reconstruction performance to be prioritised for each wavelength. Therefore optimising for overall reconstruction performance ability and stability will give us the best-performing metasurface design for a multispectral system. The acceptable maximum condition number for a polarimetry system depends on the required precision of the polarisation measurement for an application. For this work we aim to have a condition number less than 10 for simulation results, corresponding to one less significant figure of precision for measurements, with condition numbers up to 15 acceptable in preliminary experimental results to demonstrate our polarisation reconstruction abilities.

\subsection{Optimisation results}
\begin{figure}[h]
\centering
  \includegraphics[width=\columnwidth]{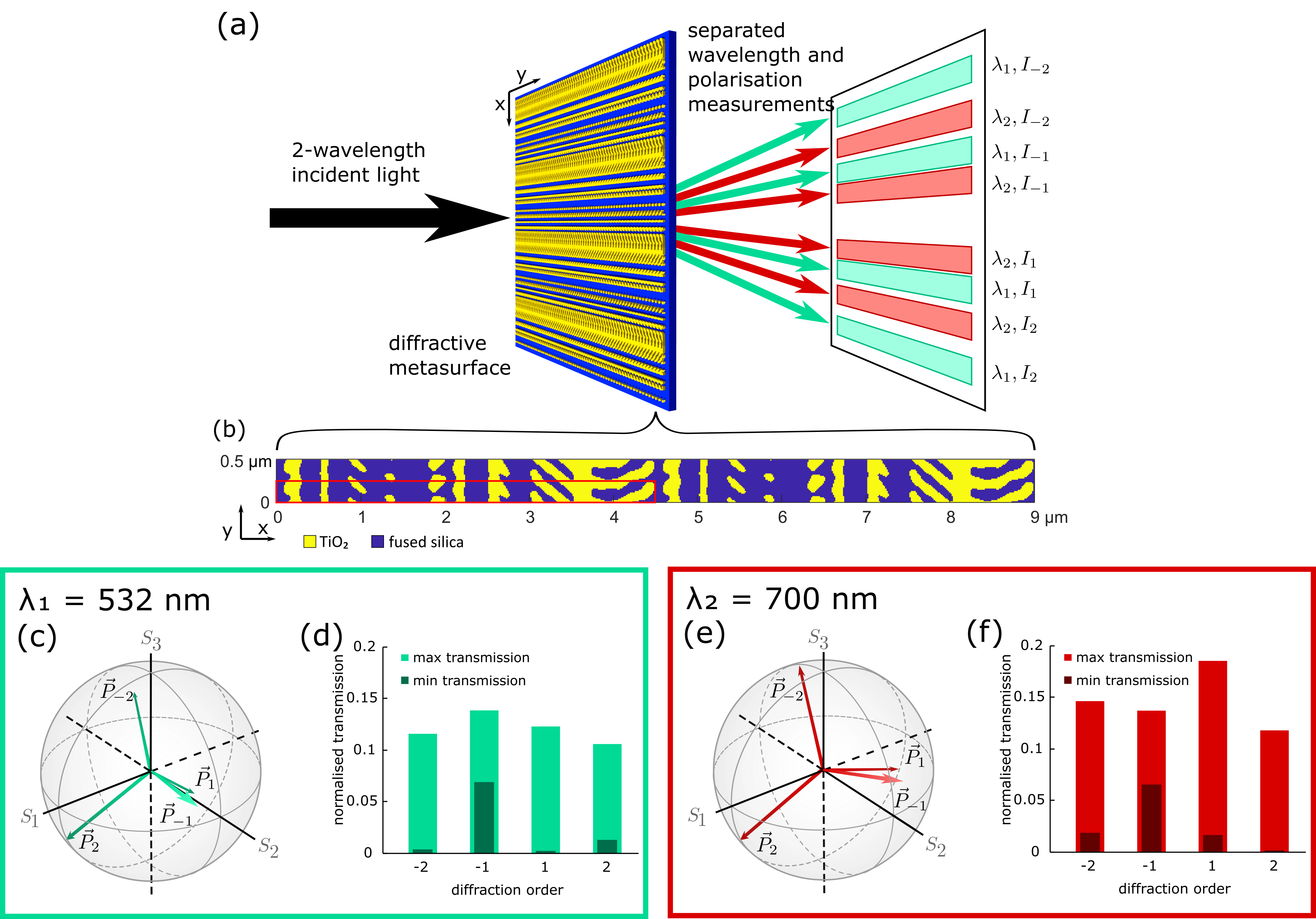}
  \caption{(a) Illustration of the metasurface behaviour. (b) The final metasurface design. A single period of size $4.5$~\unit{\micro\metre}~$\times$~$0.25$~\unit{\micro\metre} is indicated in red. (c-f)~Simulation results for incident wavelengths of (c,d)~532~\unit{\nano\metre}  and (e,f)~700~\unit{\nano\metre}. (c,e) The partial polariser associated with each intensity measurement is plotted on a Stokes sphere, where the direction indicates the polarisation. (d,f) The transmission of the ideal and orthogonal polarisation states for each diffraction order. The contrast in the transmission indicates the polarisation extinction of each diffraction order. }
  \label{fig:basicresults}
\end{figure}

Following the optimisation process, the metasurface structure with the highest multispectral polarimetry performance is shown in Fig.~\ref{fig:basicresults}(b).

Each diffraction order $n$ acts as a partial polariser, with a polarisation state $\Vec{P}_n$ and a polarisation extinction percentage between the polarisation analysis state and the orthogonal polarisation state. The polarisation analysis states corresponding to each diffraction order can be represented as a vector on the Stokes sphere to visualise the robustness of the span of these vectors in Stokes space.

To find the polarisation analysis states, we follow the method derived by Rubin \textit{et al}.\cite{Rubin2022}. The total polarisation behaviour of each diffraction order $n$, including overall transmission and polarisation diattenuation, can be represented by a vector $\Vec{D}_n=[D_{n,0},\;D_{n,1},\;D_{n,2},\;D_{n,3}]^T$, such that it is related to the instrument matrix described in Section \ref{sec:pol} as:

\begin{equation}
    \mqty[\Vec{D}_{-2}^T \\ \Vec{D}_{-1}^T \\ \Vec{D}_{1}^T \\ \Vec{D}_{2}^T]\Vec{S}=\mathbf{M}\Vec{S}=\Vec{I}.
\end{equation}
Here $D_{n,0}$ is the fraction of incident light transmitted to the $n$th diffraction order, averaged over all possible incident polarisations, and $D_{n,1},\;D_{n,2},\;D_{n,3}$ describe the polarisation behaviour. To get the polarisation analysis state $\Vec{P}_n$, we normalise the polarisation component of $\Vec{D}_n$ to give a normalised Stokes polarisation state:

\begin{equation}
    \Vec{P}_n = \frac{1}{\sqrt{D_{n,1}^2+D_{n,2}^2+D_{n,3}^2}}[D_{n,1},\;D_{n,2},\;D_{n,3}]^T.
\end{equation}
Plotting $\Vec{P}_n$ on a Stokes sphere allows the polarisation state of a diffraction order to be visualised in relation to the other diffraction orders, allowing us to ensure the analysing polarisations collectively span Stokes polarisation space, shown in Figs.~\ref{fig:basicresults}(c) and~(e). The polarisation analysis states spanning Stokes polarisation space ensure that they can be used to reconstruct an unknown polarisation with minimal error amplification.

The qualitative span of the analysis polarisation states across the Stokes space is verified by the condition number of the instrument matrix at both wavelengths, with condition numbers of 3.69 at 700~\unit{\nano\metre} and 3.29 at 532~\unit{\nano\metre}, comparable to the minimum condition number and indicating the states span the Stokes space well, and errors in the system are not significantly amplified during reconstruction. 

To determine the polarisation extinction of each diffraction order to polarisation, we calculate the contrast between the transmission of the analysis polarisation state $\Vec{P}$ and its orthogonal state, which give a maximum transmission $I_{n,\mathrm{max}}$ and a minimum transmission $I_{n,\mathrm{min}}$ shown in Figs.~\ref{fig:basicresults}(d) and~(f). Calculating the contrast in transmission as a percenatge using $(I_{n,\mathrm{max}}-I_{n,\mathrm{min}})/I_{n,\mathrm{max}}$, we find the polarisation extinction between states is at least $51\%$ across both wavelengths, with an average contrast of $82\%$, ensuring the extinction between states is large enough to distinguish different polarisations.

In addition to the polarisation reconstruction performance at the designed wavelengths and at normal incident angle, it is important to also interrogate the change in performance over a range of incident angles and wavelengths to find the imaging field-of-view and spectral bandwidth that our design can operate over.

\subsection{Angular extent of the field-of-view}

To ensure the metasurface is capable of forming images matching the requirements of a given application and imaging system, it is important to interrogate the full angular field-of-view over which the polarimetry performance of the metasurface is maintained.

We focus on the change in reconstruction performance of the collective set of measurements rather than the change in individual measurements; as each incident angle corresponds to a different spatial position on the detector, the reconstruction can be performed accounting for the changing matrix behaviour, using pixel-wise instrument matrices if necessary.

\begin{figure}[h]
\centering
  \includegraphics[width=\columnwidth]{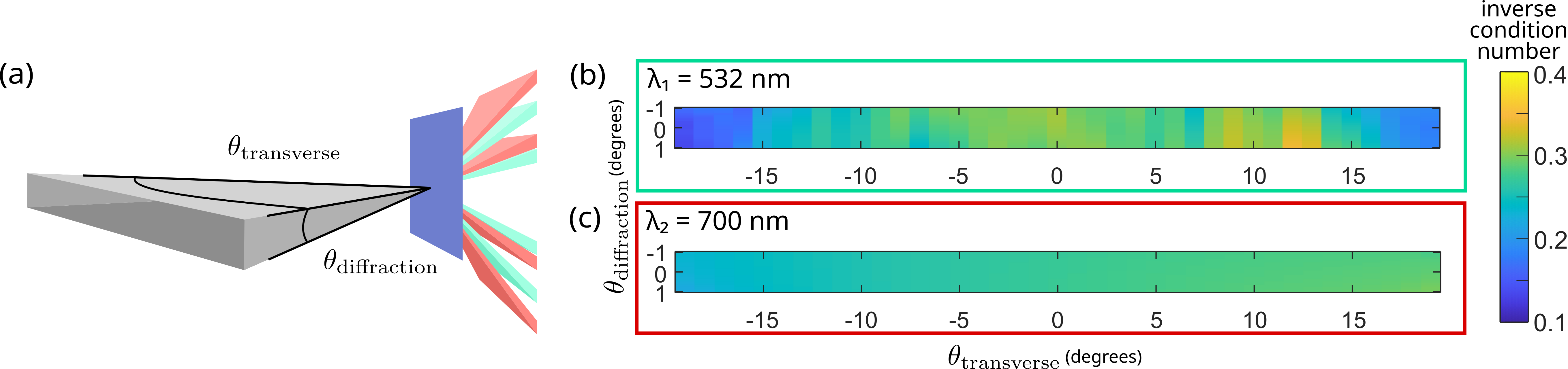}
  \caption{(a)~The metasurface is simulated for various angles of incidence covering a field-of-view (FoV) as illustrated in the sketch. (b,c)~The simulated performance across the FoV is given for (b)~532~nm and (c)~700~nm incident wavelengths, showing a condition number of less than 8 across the whole FoV. The inverse condition number, $1/\mathrm{cond}$, is used to show the change in condition number across the FoV graphically.}
  \label{fig:fov}
\end{figure}

We simulate our metasurface over a range of incident angles to scan a two-dimensional field-of-view (FoV) of $\theta_{\text{transverse}} \times \theta_{\text{diffraction}} = 36^\circ\times2^\circ$, illustrated in Fig.~\ref{fig:fov}(a), and record the polarisation performance for each angle. The tested FoV is chosen such that different diffraction orders do not overlap, and the angular width of the swath is the same as the angular extent of the measured diffraction orders such that the complete set of measurements forms a square footprint on a sensor.

The results of the angular simulations are shown in Figs.~\ref{fig:fov}(b) and~(c) for 532~nm and 700~nm incident light, respectively. The full FoV maintains a condition number of less than 8 for both wavelengths, ensuring polarisation reconstruction is robust under the angular incidence conditions of imaging.

\subsection{Spectral bandwidth \label{sec:sb}}

Another key metric for the performance of the polarisation imaging metasurface is spectral bandwidth, as the metasurface needs to maintain performance over a wide enough bandwidth such that enough light is passed through the system to achieve the necessary intensity sensitivity. We focus on determining the maximum achievable bandwidth in order to generalise our result for broader applications. There are two factors that can limit the spectral performance; the polarisation reconstruction performance at different wavelengths, and image smearing in the diffractive direction due to the Point Spread Function of the system increasing with larger spectral bandwidths.

\begin{figure}[h]
\centering
  \includegraphics[width=\columnwidth]{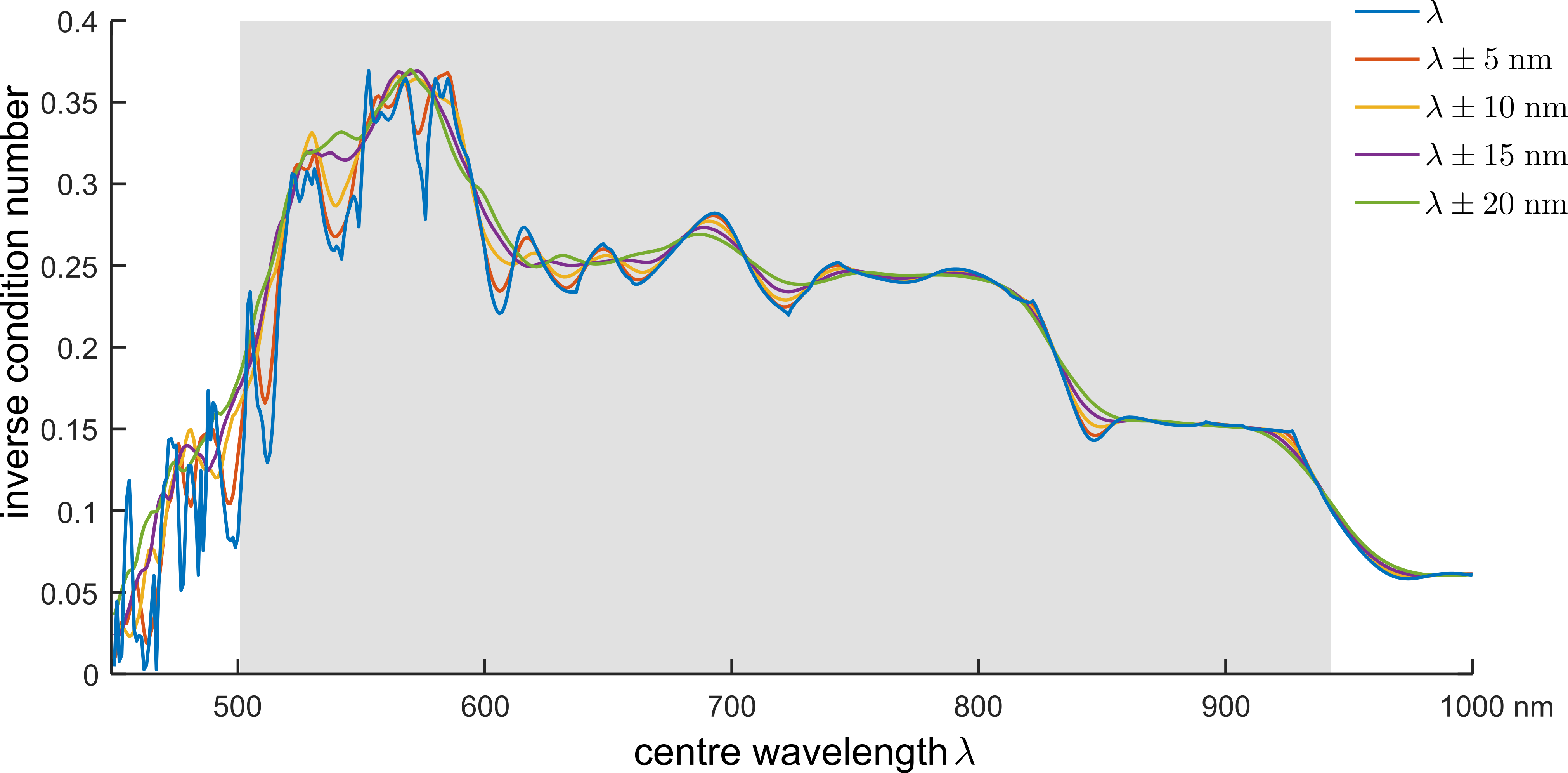}
  \caption{Performance of the metasurface simulated across a monochromatic spectrum and for several spectral bandwidths. The wavelength range that satisfies a maximum acceptable condition number of 10 is indicated in grey for the monochromatic case. The inverse condition number, $1/\mathrm{cond}$, is used to show the change in condition number across the bandwidth.}
  \label{fig:spectral}
\end{figure}

The performance of the metasurface is simulated for a range of wavelengths, shown in Fig.~\ref{fig:spectral}. When taking a maximum condition number of 10 as the threshold of performance, the range in which any monochromatic signal would sufficiently perform polarimetry is 500~\unit{\nano\metre}-942~\unit{\nano\metre}. As significant spatial overlap is expected within a diffraction order for similar wavelengths, we are unable to perform completely independent polarisation reconstructions for each wavelength in a spectral bandwidth. Therefore, we also demonstrate that the metasurface polarisation reconstruction can be calibrated for the average response for a select set of bandwidths of $\pm\delta\lambda$ around a central wavelength of $\lambda$. We show the performance of the metasurface is robust to a bandwidth of $\pm20$~\unit{\nano\metre}, provided the spectral response has a constant intensity across the bandwidth. For cases where the intensity of the spectral response varies significantly with wavelength such that the average response is not suitable for reconstruction, we have explored the variation in the metasurface instrument matrix in Section~S1  of the supplementary.

Finally, we present analysis of the image smearing due to the diffraction of a bandwidth of light, which primarily affects the resolution in the $x$-direction of the imaging. Assuming an idealised one-dimensional imaging swath, we calculate the maximum bandwidth that can be imaged before the diffraction orders overlap to be $\pm 39.2$~\unit{\nano\metre} around the designed wavelengths. This value acts as a FoM for the image resolution due to diffractive smearing; to achieve $N$ imaging samples in the diffractive direction, the operating bandwidth is a fraction of the maximum bandwidth $\pm 39.2/N$~nm.

We find that for our metasurface design and envisioned application the primary limiter is balancing the required resolution in the diffraction direction with the width of the desired spectral bandwidth, provided the intensity of the spectral response is approximately constant across our chosen spectral bandwidth.

\section{Experimental results}

\begin{figure}[h]
\centering
  \includegraphics[width=\columnwidth]{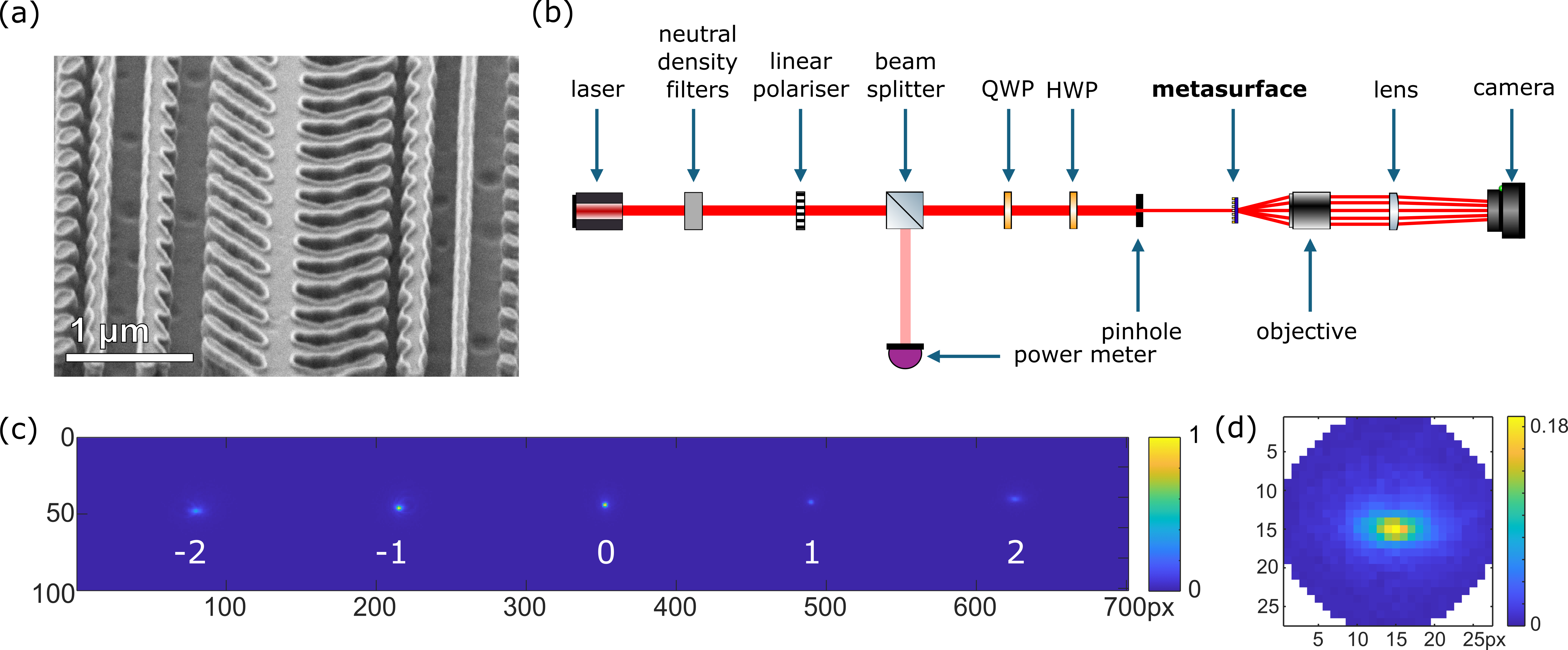}
  \caption{(a) SEM image of the fabricated metasurface. (b) Schematic of the experimental set up. Incident polarisations are prepared with a fixed linear polariser and rotating half (HWP) and quarter (QWP) waveplates, propagated through the metasurface, and captured with appropriate optics onto a camera. When necessary, the quarter waveplate was removed to prepare strictly linear polarisation states. (c) Normalised experimental camera image for one set of measurements at a monochromatic test wavelength, in this case $532$~nm, demonstrating the characteristic output diffraction pattern. The centre diffraction order is present experimentally, but discarded for analysis of the results. (d) Example region of interest for integration of the measurement for the $-2$ diffraction order, taken from the normalised camera image with colour bar rescaled to show the capture clearly.}
  \label{fig:calibration}
\end{figure}

The metasurface was fabricated using electron beam lithography using a cell projection method to efficiently fabricate several large-area metasurfaces\cite{Haedrich2022}. During preparation for fabrication it was determined necessary to include a 100~nm Al\textsubscript{2}O\textsubscript{3} layer between the TiO\textsubscript{2} layer and the fused silica substrate to ensure the TiO\textsubscript{2} is etched correctly; the optical effect of a 100~nm Al\textsubscript{2}O\textsubscript{3} layer was determined to only have a minor impact on the polarimetry performance at normal incidence such that it was not necessary to modify the structure of the patterned layer, see Supplementary~S2. The full fabrication details of the samples used in this work can be found in Section~\ref{sec:fab}. A scanning electron microscopy image of the fabricated structure shown in Fig.~\ref{fig:calibration}(a), and the performance of this metasurface is evaluated through calibration and testing at two incident wavelengths.

The characterisation and testing of the metasurface was performed using the set-up illustrated in Fig.~\ref{fig:calibration}(b); input elliptical polarisation states are prepared using a linear polariser and rotating quarter and half waveplates. The prepared incident states are transmitted through the metasurface, and the $\pm 1$ and $\pm 2$ diffraction orders captured with a $20\times$ objective with $\mathrm{NA} = 0.4$, exceeding the minimum capture $\mathrm{NA}$ for our design calculated in Section~\ref{sec:framework}. The diffraction orders are imaged with a camera sensor, shown in Fig.~\ref{fig:calibration}(c), with the power of the laser recorded simultaneously for normalisation. The camera image is integrated over a region of interest, as pictured in Fig.~\ref{fig:calibration}(d), for each of the $\pm 1$ and $\pm 2$ diffraction orders to give the full intensity measurement for an input polarisation state. Additional information on the preparation and measurement of test states can be found in Supplementary~S3.

The behaviour of the metasurface is calculated by preparing a set of known input states, recording the output intensity pattern, and calculating the instrument matrix that describes the relationship between the input and output states, as per Eq.~(\ref{eq:MMat}). A series of test states, measured independently of the calibration states, are reconstructed using the calculated instrument matrix to evaluate the errors in the reconstruction.

\subsection{Multispectral polarisation performance}

 Due to the availability of laser sources, we choose to perform our experimental evaluation at $532$~nm and $637$~nm. These wavelengths still effectively demonstrate the multispectral capabilities of the fabricated metasurface, as the simulated spectral performance of our design in Section \ref{sec:sb} shows we expect similar polarimetric capabilities at 637~nm and 700~nm incident wavelengths.

 \begin{figure}[h]
\centering
  \includegraphics[width=\columnwidth]{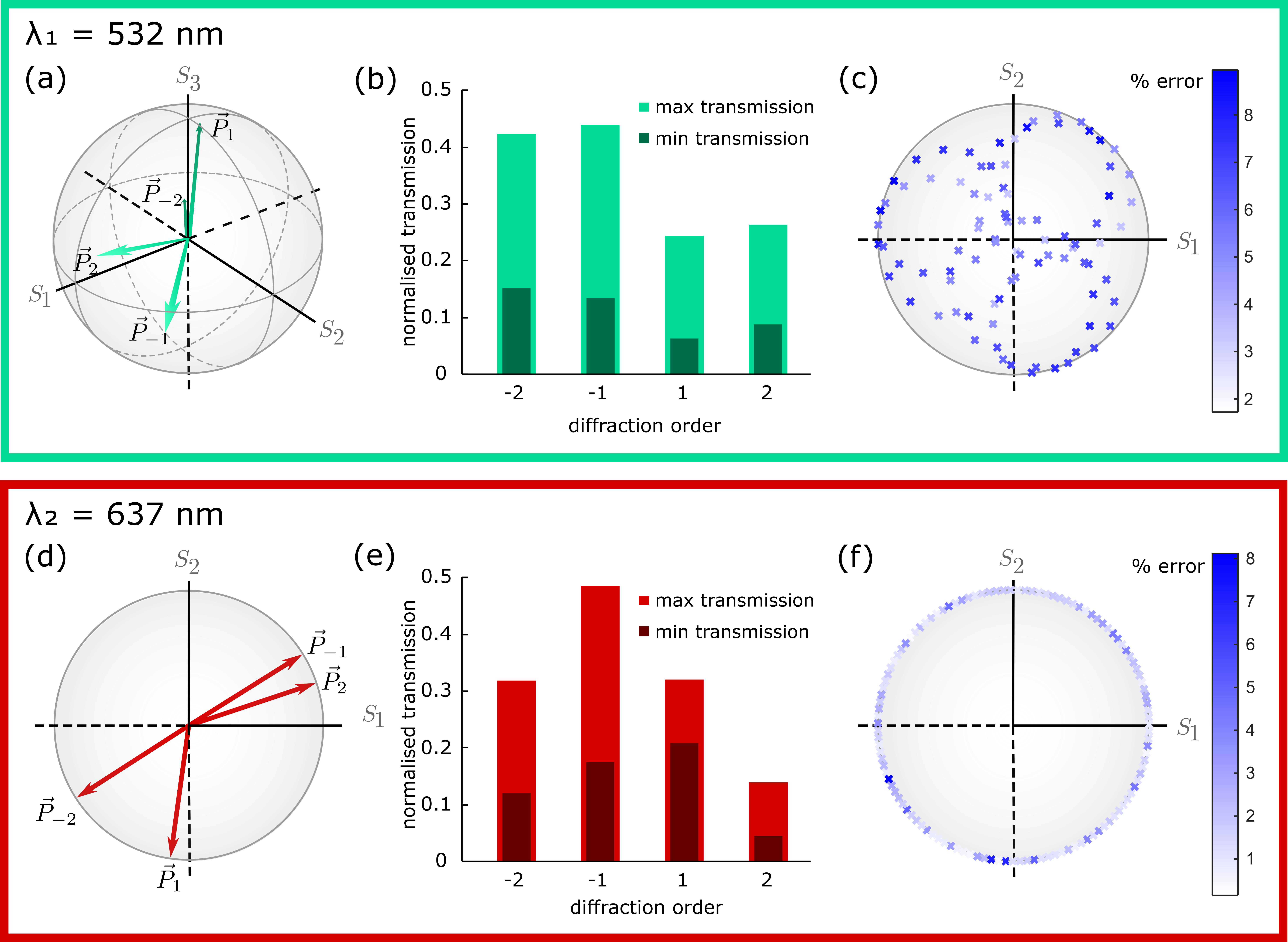}
  \caption{Experimental results of the metasurface calibration for the two wavelengths $\lambda_1$, $\lambda_2$ used in the experiment, showing (a-c)~full-Stokes polarimetry at $\lambda_1 = 532$~nm and (d-f)~linear polarimetry at $\lambda_2 = 637$~nm. (a) Full stokes and (d) linear polarisation analysis states of the experimental device at the respective wavelengths. (b,e)~The transmission of the polarisation analysis state and its orthogonal state for each diffraction order, demonstrating the polarisation extinction of the metasurface. The transmission is normalised by the calibration measurement with the highest total transmission. (c,f)~Demonstration of the reconstruction ability using (c)~generally elliptical and (f)~linearly fully polarised test states, shown in the $S_1$-$S_2$ plane of Stokes space.
  A cross denotes an input polarisation, with the colour of the cross indicating the root mean square error between the true and reconstructed test state as a percentage.}
  \label{fig:experiment}
\end{figure}

 The exact behaviour of the metasurface diverges from the optimised design, which we attribute to fabrication variations that are unable to be accounted for in a single fabrication cycle. However, the nature of our calibration process ensures we do not rely on our fabrication precisely matching the optimised design and is robust to such structural variations. We therefore focus on evaluating our metasurface per the actual polarisation response of the specific fabrication sample used, demonstrating the robustness of our metasurface platform paired with our calibration method for polarimetry.
 
 The performance at $532$~nm is shown to be suitable for reconstruction of the full-Stokes polarisation state. The analyser states associated with each diffraction order, Fig.~\ref{fig:experiment}(a), span the Stokes space with a condition number of $14$. The errors in the system are not amplified excessively and polarisation reconstruction is still achievable with the experimental polarisation analysis states, despite the divergence in metasurface behaviour and performance from the design simulation. The polarisation extinction of each diffraction order is at minimum $64.4\%$ of the transmission, shown in Fig.~\ref{fig:experiment}(b), ensuring different incident polarisations can be distinguished. 

 The success of the reconstruction of a test polarisation state is given by the difference between the prepared polarisation state and the reconstructed state. To quantify the difference between two polarisation states, $\Vec{P}_{a}$ and $\Vec{P}_{b}$, we use the Root Mean Squared Error (RMSE). Using $\delta \Vec{P}=\Vec{P}_{a}-\Vec{P}_{b}=(\delta P_{0}, \delta P_{1}, \delta P_{2}, \delta P_{3})$, the $\mathrm{RMSE}$ is calculated as:
\begin{equation}
    \mathrm{RMSE}(\Vec{P}_{a},\Vec{P}_{b}) = \sqrt{\frac{1}{4} \sum_{i=0}^3\abs{\delta P_{i}}^2}.
    \label{eq:rmse}
\end{equation}
The overall success of reconstructing the test polarisation states is given by the average RMSE over each test state reconstruction.
 
 Reconstruction of test states using our calculated instrument matrix and the maximum likelihood method are successful, with an average RMSE of $4.43\%$ and the largest error during testing being $8.51\%$. The reconstruction errors between the true and measured test state are shown in Fig.~\ref{fig:experiment}(c). Therefore, the metasurface meets the performance metrics required for full-Stokes polarimetry at an incident wavelength of $532$~nm.

 Experimentally, the metasurface sample is unable to perform full-Stokes polarimetry at $637$~nm, see Supplementary~S4. However, we are still able to demonstrate polarisation-selective behaviour by calibrating the metasurface for exclusively linear polarimetry. This calibration is performed by preparing linear polarisation states $\Vec{S}=(S_0,\;S_1,\;S_2)$ and measuring the output intensity pattern, and then calculating the instrument matrix $\mathbf{M}$ analogously to the full-Stokes calibration by solving Eq.~(\ref{eq:MMat}). The linear instrument matrix is a $4 \times 3$ matrix due to the four measurements forming an overdetermined system, and can be analysed similarly to the full-Stokes case; the condition number describes the span of the polarisation analysis vectors across linear Stokes space, and the polarisation extinction is calculated from the maximum and minimum transmission.
 
 The condition number for the linear-only calibration of the metasurface is $15$, with the analysis vectors spanning the linear Stokes space as shown in Fig.~\ref{fig:experiment}(d), sufficient for polarisation reconstruction without prohibitively amplified system errors. The polarisation extinction of some analysis states is lower for the linear case, Fig.~\ref{fig:experiment}(e), but is above 60\% for the -2,~-1, and the +2 diffraction orders, and still 35\% for the +1 order, ensuring we can distinguish linear polarisation. 
 
 The RMSE for the linear case is similar to Eq.~(\ref{eq:rmse}), just for three components we use $\delta \Vec{P}=\Vec{P}_{a}-\Vec{P}_{b}=(\delta P_{0}, \delta P_{1}, \delta P_{2})$, and the $\mathrm{RMSE}$ is calculated as:
\begin{equation}
    \mathrm{RMSE}(\Vec{P}_{a},\Vec{P}_{b}) = \sqrt{\frac{1}{3} \sum_{i=0}^2\abs{\delta P_{i}}^2}.
\end{equation}

Linear test polarisation states are reconstructed with an average RMSE of $2.31\%$ and a highest test error of $8.12\%$, Fig.~\ref{fig:experiment}(f), demonstrating successful reconstruction of linear test states. 
\section{Discussion}

We have presented a metasurface design capable of performing multispectral polarisation imaging within a single metasurface structure, avoiding filtering and imaging offsets that can be detrimental to imaging. We evaluated the spatial and spectral performance of our metasurface design to demonstrate the extent of performance, verifying that the metasurface could be used in realistic imaging systems. Furthermore, we fabricated our designed metasurface and performed preliminary calibration at two wavelengths, achieving full-Stokes polarimetry at $532$~nm and linear polarimetry at $637$~nm.

Metasurface-based full-Stokes polarisation imagery drastically reduces the size and optical complexity of a system compared to conventional polarisation optics. This enables the addition of polarimetry to a broad range of applications, beyond the agricultural surveying application motivating this work, where multispectral polarisation information could greatly enhance imaging.

\section{Fabrication Methods\label{sec:fab}}
The samples were fabricated on 100~mm diameter, 0.5~mm thick fused-silica substrates. After cleaning in piranha solution, a 100~nm Al\textsubscript{2}O\textsubscript{3} layer and a 319~nm TiO\textsubscript{2} layer were deposited consecutively without breaking vacuum by ion-beam sputter deposition using an Ionfab 300LC system (Oxford Instruments Plasma Technology) at an ion energy of 1300~eV and a beam current of 350~mA. Prior to deposition, the substrate surface was pre-etched for 15~s using 400~eV Ar ions. Subsequently, a 50~nm Cr hard-mask layer was deposited under identical ion-beam conditions. After vapor-phase HMDS treatment in a Sawatec 200/300 system, a 100~nm thick layer of the electron-beam resist OEBR-CAN038 AE 2.0CP (Tokyo Ohka Kogyo Co., Ltd.) was applied by spin coating. The pattern was exposed using a Vistec 350OS electron-beam writer equipped with character-projection apertures\cite{Haedrich2022}, developed manually in OPD4262 for 30~s, and thoroughly rinsed with deionized water. The resist pattern was transferred into the Cr layer by ion-beam sputter etching at 400~eV, with the etch endpoint determined by optical emission spectroscopy. The remaining resist was removed by oxygen-plasma ashing for 5~min in a Sentech SI 591 system. The pattern was subsequently transferred into the TiO\textsubscript{2} layer by inductively coupled plasma etching in a Sentech SI 500 C system using SF\textsubscript{6}/O\textsubscript{2}/Ar gas flows of 20/5/6~sccm and an ICP power of 700~W. Finally, the Cr hard mask was removed in the Sentech SI 591 using Cl\textsubscript{2} and O\textsubscript{2} gas flows of 50 and 10~sccm, respectively, before the wafer was diced into individual samples.

\section*{Data Availability}

All data generated or analysed during this study are visually presented this published article and its supplementary is available from the corresponding author upon reasonable request.
\section*{Code Availability}

The underlying code created for this study is not publicly available, but may be made available to qualified researchers on reasonable request from the corresponding author. 

\section*{Acknowledgements}
We acknowledge the funding support of the Australian Research Council (CE200100010) and the Deutsche Forschungsgemeinschaft (DFG, German Research Foundation) through the International Research Training Group (IRTG) 2675 “Meta-Active” (project number 437527638),

The sample fabrication within this work was carried out by the microstructure technology team at IAP Jena. The authors would like to thank them for providing the fabrication facilities, carrying out processes and providing support.

\section*{Author Contributions}

S.E.D. contributed to the methodology for the simulation and experiment, performed the topology optimisations for the metasurface design, built the experimental characterisation set-up and acquired experimental data, performed analysis of both the simulation and experimental data, and wrote the manuscript. N.L. provided the topology optimisation methodology and the core implementation of the topology optimisation for simulations. J.M. conceptualised the experimental characterisation set-up, and contributed to the analysis code for the experimental data. B.L. and G.Q.N. contributed to refinement of the experimental characterisation set-up and full-Stokes polarisation data acquisition. B.L. also acquired data for linear polarisation characterisation. T.S. provided fabrication information for the optimisation process, and fabricated the metasurface samples. B.L. R.S., D.N.N., F.E., and A.A.S. contributed to the review and preparation of the manuscript. R.S., D.N.N., F.E., and A.A.S. supervised and conceptualised this work. All authors read and approved the final manuscript.

\section*{Competing Interests}

All authors declare no financial or non-financial competing interests.


\bibliography{visrefs}   
\bibliographystyle{spiejour}   


\end{spacing}
\end{document}